\documentclass[prl, showpacs,twocolumn]{revtex4} \usepackage{latexsym}
\usepackage{amsfonts} \usepackage{amsmath,revsymb}

\newcommand{\bs}{\boldsymbol}
\begin{document}

\title{Loop prolongations and three-cocycles in simulated magnetic fields from rotating reference frames}

\author{W.H.~Klink$^{a}$ and S.~Wickramasekara$^{a, b}$}
\affiliation{$^{a}$Department of Physics and Astronomy, University of Iowa, Iowa City, IA 52242\\
$^{b}$Department of Physics, Grinnell College, Grinnell, IA 50112}

\begin{abstract}
We show that the Wigner-Bargmann program of grounding non-relativistic quantum mechanics in the unitary projective representations of the Galilei group can be extended 
to include all non-inertial reference frames. The key concept is the \emph{Galilean line group}, the group of transformations that ties together all accelerating reference frames, and its 
representations. These representations are constructed under the natural constraint that they reduce to the well-known unitary, projective representations of the Galilei group when the 
transformations are restricted to inertial reference frames. This constraint can be accommodated only for a class of representations with a sufficiently rich cocycle structure. 
Unlike the projective representations of the Galilei group, these cocycle representations of the Galilean line group do not correspond to central extensions of the group. Rather, they 
correspond to a class of non-associative extensions, known as \emph{loop prolongations}, that are determined by three-cocycles. As an application, we show that the phase shifts
due to the rotation of the earth that have been observed in neutron interferometry experiments and the rotational effects that lead to simulated magnetic fields in optical lattices 
can be rigorously derived from the representations of the loop prolongations of the Galilean line group.
\end{abstract}
\maketitle
\noindent\emph{Introduction:} Following the very interesting experiment of Werner, Staudenmann and Colella \cite{werner} that measured an interference effect in the neutron wave function due to the rotation 
of the earth, there have been several attempts to derive the non-inertial effects of a rotating reference frames in quantum mechanics. The first among these attempts was by Sakurai \cite{sakurai} who used the 
similarity between the Lorentz force law $\bs{F}=e\bs{v}\times\bs{B}$ and  the Coriolis force $\bs{F}=2m\bs{v}\times\Omega$ to calculate the phase shift in the neutron interferometry 
experiment of \cite{werner}.  Later, Mashhoon \cite{mashhoon} used what he called a ``simple, yet tentative, extension of the hypothesis of locality" to obtain the same phase shift. 
Anandan critiqued this reasoning \cite{anandan1} and later Anandan and Suzuki \cite{anandan2} studied the problem by drawing on the analogy between the 
Galilean transformation properties of the Schroedinger equation and the minimal coupling of a $U(1)$-gauge connection. All of these approaches are heuristic, essentially relying on inspired appeal to analogy rather than 
being grounded on first principles. The purpose of this paper is to present a formulation of quantum mechanics that holds in all non-inertial reference frames and show that the Coriolis effect of rotating reference frames leading to the 
observed phase shift naturally follows from this formulation.  

On the experimental side, the analogy between the Lorentz and Coriolis forces continues to be explored. 
The key observation is that the effect of the Coriolis force on a massive particle, be it charged or neutral, 
is identical to the effect of an equivalent magnetic field on a charged particle. In this regard, a suitable rotating frame can be used to overcome limitations that arise 
from the charge neutrality of atoms in cold atom experiments.  In fact, several remarkable experiments  \cite{zwierlein,schweikhard,madison,abo-shaeer,lin} have shown 
the appearance of  vortices in rotating atomic gases and Bose-Einstein condensates, a property generally attributed to superfluids and superconductors in magnetic fields. 
Simulated magnetic fields generated by the Coriolis effect have also been demonstrated by the behavior of rotating atomic systems in the 
fractional Hall effect  \cite{gemelke}. The formalism presented here provides a theoretical framework for understanding these varied experimental results. \\

\noindent\emph{Galilean line group:} Consider spacetime transformations
\begin{eqnarray}
\bs{x}'&=&R(t)\bs{x}+\bs{a}(t)\nonumber\\
 t'&=&t+b\label{1}
\end{eqnarray}
where rotations $R$ and space translations $\bs{a}$ are functions of time. Together, they define the \emph{Euclidean line group} ${\cal E}(3)$, 
the infinite dimensional group of functions on the real line taking values in the three-dimensional Euclidean group \cite{klink1}.  When acting on a spacetime point $(\bs{x}, t)$, $R$ and 
$\bs{a}$ get evaluated at $t$, leading to \eqref{1}. It follows from \eqref{1} that the transformations 
$(R,\bs{a},b)$ form a group: 
\begin{equation}
(R_2,\bs{a}_2,b_2)(R_1,\bs{a}_1,b_1)=((\Lambda_{b_1}R_2)R_1,\Lambda_{b_1}\bs{a}_2+(\Lambda_{b_1}R_2)\bs{a}_1,b_2+b_1)\label{3}
\end{equation}
where $\Lambda_b$ is the shift operator $(\Lambda_bf)(t)=f(t+b)$. It accounts for the fact that in a successive application of \eqref{1}, 
the group elements $R_2$ and $\bs{a}_2$ get evaluated at $t+b_1$, whereas $R_1$ and $\bs{a}_1$ get evaluated at $t$. 
We call the group defined by \eqref{3} the \emph{Galilean line group} and denote it by ${\cal G}$. 
It is straightforward to verify that $\Lambda_b$ is an automorphism on ${\cal E}(3)$ and that  
${\cal G}$ is the semidirect product of ${\cal E}(3)$ and $\mathbb{R}$ with respect to $\Lambda_b$. 

For $R$ constant and $\bs{a}(t)$ of the form $\bs{a}(t)=\bs{a}^{(0)}+\bs{v}t$, \eqref{1} reduces to the usual Galilean transformations and the 
group law \eqref{3} becomes the composition rule for the Galilei group. Hence, the Galilean line group contains the Galilei group as a subgroup. More generally, 
$R$ and $\bs{a}$ are arbitrary functions of time and, as such, they implement transformations among all rotationally and linearly accelerating reference frames. For instance, 
$\bs{a}(t)=\bs{a}^{(0)}+\bs{v}t+\frac{1}{2}\bs{a}^{(2)}t^2$ furnishes transformation to a uniformly accelerating reference frame. 
Therefore, ${\cal G}$ is the group of transformations that ties together all reference frames of a Galilean spacetime.  Extending the Wigner- Bargmann program, 
our principle claim is that a Galilean quantum theory that holds in non-inertial reference frames should be grounded in unitary representations of the Galilean line group. 
We will show below that the  quantum mechanical effects of rotational reference frames discussed  in the introduction, in particular the simulation of magnetic effects,  can be obtained from these representations. 

\noindent\emph{Loop prolongations:}
If a representation of the Galilean line group is to describe a particle, it must contain as a subrepresentation a unitary, irreducible, projective representation 
of the Galilei group because it is these representations that describe particles in Galilean quantum mechanics of inertial reference frames \cite{bargmann,leblond}. 
Recall that a unitary projective representation is one in which the homomorphism property holds only up to a phase, called a \emph{two-cocycle}: 
 \begin{equation}
 \hat{U}(g_2)\hat{U}(g_1)=e^{i\omega(g_2,g_1)}\hat{U}(g_2g_1).\label{5}
 \end{equation}
The associativity requirement for the operators $\hat{U}(g)$  leads to a constraint on the two-cocycle:  
\begin{equation}
\omega(g_3,g_2g_1)+\omega(g_2,g_1)
 -\omega(g_3g_2,g_1)-\omega(g_3,g_2)=0. \label{7}
\end{equation}
Bargmann showed that a projective representation of a group $G$ is equivalent to a true representation of its central extension $\bar{G}=\{\bar{g}:\ \bar{g}=(\varphi,g), g\in G, \varphi\in\mathbb{R}\}$ by $\mathbb{R}$, defined 
by the composition rule 
\begin{equation}
(\varphi_2,g_2)(\varphi_1,g_1)=(\varphi_2+\varphi_1+\frac{1}{m}\omega(g_2,g_1), g_2g_1),\label{11}
\end{equation}
where $m$ is an arbitrary non-zero number. That \eqref{11} is associative is ensured by \eqref{7}. 
The  true representation of $\bar{G}$ that is equivalent to \eqref{5} is defined by $\hat{U}(\bar{g}):=e^{im\varphi}\hat{U}(g)$. It follows  from \eqref{5} and \eqref{11} that 
$\hat{U}(\bar{g}_2)\hat{U}(\bar{g}_1)=\hat{U}(\bar{g}_2\bar{g}_1)$. 

The unitary irreducible representations of the centrally extended Galilei group can be constructed by 
the Wigner-Mackey method of induced representations~\cite{bargmann,leblond}. For the spin-zero case, to which we limit ourselves in order to avoid inessential complications, the representation can be defined 
by the transformation formula 
\begin{equation}
\hat{U}(\bar{g})|\bs{q}\rangle=e^{im(\varphi+\bs{q}'\cdot\bs{a}^{(0)}-\frac{1}{2}\bs{v}\cdot\bs{a}^{(0)}+\frac{1}{2}\bs{q}'^2b)}|\bs{q}'\rangle\label{13}
\end{equation}
where $\bs{q}'=R\bs{q}+\bs{v}$, the Galilean transformation formula for velocity. (We label states by velocity rather than momentum.)  Without loss of generality \cite{leblond}, we have set the invariant internal energy to be zero. Now, 
substituting \eqref{13} in \eqref{5}, we deduce the two-cocycle for the Galilei group: 
\begin{equation}
\omega(g_2, g_1)=\frac{m}{2}\left(\bs{a}^{(0)}_2\cdot R^{(0)}_2\bs{v}_1-\bs{v}_2\cdot R_2^{(0)}\bs{a}_1^{(0)}+b_1\bs{v}_2\cdot R^{(0)}_2\bs{v}_1\right)\label{9}
\end{equation}
where  the superscript $(0)$ indicates the time independent nature of rotations and spatial translations of the Galilei group (zeroth order Taylor coefficients of $R$ and $\bs{a}$ of 
${\cal G}$). 

Our task is to construct representations of the Galilean line group under the constraint that they reduce to \eqref{13}. 
By the Bargmann correspondence between projective representations and central extensions, this reduction requirement can be achieved by constructing an embedding of a given central extension of the Galilei group 
in an extension of the Galilean line group. Here we face a difficulty: while the Galilean line group naturally contains the Galilei group as a subgroup, as shown in \cite{macgregor}, 
\emph{there exist no group extensions, central or non-central, of the Galilean line group that fulfills this embedding requirement}. Three-cocycles  arise out of this limitation. 

Stated another way, there exist no two-cocycles of the Galilean line group that reduce to the Galilean two-cocycle (7).  This conclusion follows from the following key properties of the cocycle structure of the Galilean line group:
\begin{enumerate}
\item Since elements of ${\cal G}$ are functions of time, any non-trivial cocycle on ${\cal G}$ must be a mapping into a group of functions of time, 
rather than just $\mathbb{R}$ as is the case  for the Galilei group. But then the automorphism $\Lambda_b$ would act non-trivially 
on these functions and the two-cocycle condition becomes
\begin{eqnarray}
(\delta\omega)(g_3,g_2,g_1)&:=&\Lambda_{b_1}\omega(g_3,g_2)+\omega(g_3g_2,g_1)-\omega(g_2,g_1)\nonumber\\
&&\ \ -\omega(g_3,g_2g_1)=0.\label{15}
\end{eqnarray}
 Therefore, any extension of the Galilean line group, if it exists, must be non-central. 
\item The reduction to the Galilean two-cocycle \eqref{9} implies that the relevant solutions to \eqref{15} must take values in ${\cal F}$, 
the  group of scalar functions on $\mathbb{R}$. 
Given such solutions to \eqref{15}, the group extensions of ${\cal G}$ by ${\cal F}$ would be defined by 
\begin{equation}
(\varphi_2,g_2)(\varphi_1,g_1)=(\Lambda_{b_1}\varphi_2+\varphi_1+\frac{1}{m}\omega(g_2,g_1), g_2g_1)\label{17}
\end{equation}
Expressions \eqref{15} and \eqref{17} are to be compared with the  ones for central extensions, \eqref{7} and \eqref{11}. 
\item 
Since \eqref{9} involves velocities, a two-cocycle of ${\cal G}$ reducing to \eqref{9} must contain 
the derivatives $\dot{\bs{a}}$ of spatial translations. This leads to an additional complication: under time dependent rotations, 
$\bs{a}$ and $\dot{\bs{a}}$ do not transform the same way owing to the inhomogeneous $\dot{R}$-term in $\frac{d}{dt}(R\bs{a})=R\dot{\bs{a}}+\dot{R}\bs{a}$. This difficulty 
is a familiar one from gauge field theories. In the present case, the trouble is algebraic: any two-cocycle of the Galilean line group taking values in ${\cal F}$ and containing derivatives of space translations 
fails to fulfill the two-cocycle condition \eqref{15}.  
\end{enumerate}
Group extensions of ${\cal G}$ by ${\cal F}$ are precluded by the failure of \eqref{15}  because it is this condition that ensures associativity of the composition rule \eqref{17}.  
However, there exist non-associative extensions of ${\cal G}$ by  ${\cal F}$ that fit very nicely into the theory of \emph{loop prolongations} 
developed by Eilenberg and MacLane \cite{em}. A loop is a set with a binary operation that fulfills  the axioms of a group, 
except for associativity and therewith also the existence of a two-sided inverse for every element (the left and right inverses may be distinct).  Further, 
given three elements $a,b,c$ of a loop $L$, there always exists a unique element $A(a,b,c)\in L$, called an \emph{associator},  
such that
\begin{equation}
a(bc)=A(a,b,c)[(ab)c].\label{21}
\end{equation}
Associators measure deviations from associativity, much like commutators measure the lack of commutativity. We will not review the general theory of \cite{em} here, but only 
mention that the construction of loop prolongations runs parallel to that of group extensions, with a little additional care to handle the complications resulting from the failure of \eqref{15}.

Since the embedding of central extensions of the Galilei group is the key requirement, the construction of a loop prolongation of ${\cal G}$ must start with a function $\omega: {\cal G}\times{\cal G}\to{\cal F}$ that 
reduces to \eqref{9}. The simplest choice is: 
\begin{equation}
\omega(g_2,g_1)=\frac{m}{2}\left((\Lambda_{b_1}\bs{a}_2)\cdot(\Lambda_{b_1}R_2)\dot{\bs{a}_1}-(\Lambda_{b_1}\dot{\bs{a}}_2)\cdot(\Lambda_{b_1}R_2){\bs{a}_1}\right).\label{23}
\end{equation}
Substituting \eqref{23} in the right hand side of \eqref{15} gives
\begin{eqnarray}
(\delta\omega)(g_3,g_2,g_1)&=&\frac{m}{2}\Lambda_{b_1}\Omega_2\cdot(\Lambda_{b_1}R_2\bs{a}_1\times\Lambda_{b_2+b_1}(R_3^T\bs{a}_3))\nonumber\\
&&-\frac{m}{2}\Lambda_{b_2+b_1}\Omega_3\cdot(\Lambda_{b_1}\bs{a}_2\times\Lambda_{b_1}R_2\bs{a}_1)\not=0\nonumber\\
\label{25}
\end{eqnarray}
where the angular velocity vector $\Omega$ is related to the time derivative of the rotation matrix by the usual formula $\Omega\times\bs{a}=\dot{R}R^T\bs{a}$. 
As mentioned above,  we see here that the time dependence of rotations violates the cocycle condition \eqref{15}. In fact, if restricted to constant rotations, 
\eqref{23} becomes a two-cocycle and leads to group extensions of the linear acceleration subgroup of the Galilean line group, albeit these extensions are non-central. 
Representations of this non-centrally extended linear acceleration group have been constructed and their role in the Galilean equivalence principle in quantum mechanics has been studied 
in \cite{klink1, macgregor}. In fact, there exist infinitely many inequivalent extensions  of the linear acceleration group, the vast majority  of which  lead to violations of the equivalence principle 
 \cite{klink3}.

It can be shown that the collection of elements $\bar{\cal G}:=\{(\varphi,g)\}$, where $g\in{\cal G}$ and $\varphi\in{\cal F}$, together with the composition rule \eqref{17}, fulfill all axioms of a loop
 prolongation of a group by an Abelian group \cite{klink2}. When elements of ${\cal G}$ are restricted 
to Galilei transformations, $\bar{\cal G}$ reduces to the central extension  of the Galilei group defined by \eqref{9}, our crucial embedding requirement. 

A direct calculation using \eqref{21} shows that the associators of $\bar{\cal G}$ are of the form
  \begin{equation}
A(\bar{g}_3,\bar{g}_2,\bar{g}_1)=(\delta\omega(g_3,g_2,g_1),e)\label{27}
\end{equation}
where $e=(I,\bs{0},0)$ is the identity element of ${\cal G}$. 
Hence, associators all belong to the Abelian subgroup ${\cal F}$ of $\bar{\cal G}$. Further, equation \eqref{25} that defines 
the associators of $\bar{\cal G}$ is a \emph{three-cocycle}. 
We call the loop prolongation $\bar{\cal G}$ defined by the three-cocycle \eqref{25} the \emph{Galilean 
line loop}. Though not common in the physics literature, three-cocycles have been studied in connection with magnetic monopoles \cite{jackiw,hou}. However, the content of the present case appears to be 
quite different from these previous studies. 

\noindent\emph{Unitary representations of the Galilean line loop:} 
The main thesis that we advocate in this letter is that quantum mechanics in non-inertial reference frames should be grounded in unitary, possibly cocycle, representations 
of the Galilean line loop. In particular, a particle should be described by a unitary, {irreducible} representation (irrep) of the Galilean line loop and multi-particle systems 
by tensor products of such irreducible representations. 

The irreps of the Galilean line loop can be constructed by the  method of induced representations. Operators $\hat{U}(\bar{g})$ furnishing the representation  can be defined 
by their action on the velocity eigenvectors $|\bs{q}\rangle$:
\begin{equation}
\hat{U}(\bar{g})|\bs{q}\rangle=e^{i\xi(\bar{g};\bs{q})}|\Lambda_{-b}\bs{q}'\rangle\label{29}
\end{equation}
where 
\begin{eqnarray}
\xi(\bar{g};\bs{q})&=&m\left(\varphi+\bs{q}'\cdot\bs{a}-\frac{1}{2}\bs{a}\cdot\dot{\bs{a}}+\frac{1}{2}(\Lambda_{-b}-1)\bs{q}'\cdot\bs{a}_{\bs{q}'}\right)\nonumber\\
\bs{a}_{\bs{q}}&=&\int dt\,\bs{q},\quad \bs{q}=\frac{d}{dt}\bs{a}_{\bs{q}}\nonumber\\
\bs{q}'&=&R\bs{q}+\dot{R}\bs{a}_{\bs{q}}+\dot{\bs{a}}=\frac{d}{dt}(R\bs{a}_{\bs{q}}+\bs{a})=\frac{d}{dt}\bs{a}_{\bs{q'}}\label{31}
\end{eqnarray}
It follows from the expression for $\bs{q}'$ that $\bs{a}_{\bs{q}}$ is our standard boost in that the operator $\hat{U}(g_{\bs{q}}), \ g_{\bs{q}}=(0,I,\bs{a}_{\bs{q}},0)$, transforms the rest 
vector $|\bs{0}\rangle$ to $|\bs{q}\rangle$. In evaluating the integral $\int dt\,\bs{q}$ to determine $\bs{a}_{\bs{q}}$, 
we choose the boundary condition that the  constant of integration be set to zero. 

The representation \eqref{29} reduces precisely to the Galiean group representation \eqref{13}. By way of this property, the $m$ in \eqref{31} acquires interpretation 
as inertial mass. Further, composing two elements, 
\begin{equation}
\hat{U}(\bar{g}_2)\hat{U}(\bar{g}_1)\mid\bs{q}\rangle=e^{i\xi_2(\bar{g}_2,\bar{g}_1;\bs{q})}\hat{U}(\bar{g}_2\bar{g}_1)\mid\bs{q}\rangle,\label{33}
\end{equation}
where 
\begin{eqnarray}
\xi_2(\bar{g}_2,\bar{g}_1;\bs{q})&:=&\xi(\bar{g}_1;\bs{q})+\xi(\bar{g}_2;\Lambda_{-b_1}\bs{q}')-\xi(\bar{g}_2\bar{g}_1;\bs{q})\nonumber\\
&=&m(\Lambda_{b_1}\Omega_2)\cdot(R_1\bs{a}_{\bs{q}}\times\bs{a}_1-(\Lambda_{b_1}R_2)\bs{a}_1\times\Lambda_{b_1}\bs{a}_2)\nonumber\\
&&+m(1-\Lambda_{b_1})\varphi_2+m(\Lambda_{-b_1}-1)\omega(g_2,g_1g_{\bs{q}})
\label{35}
\end{eqnarray}
It is evident from \eqref{35} that the representation structure of the Galilean line loop is more intricate than that of projective representations of groups. In the latter case, once the central extension has been 
taken, the equivalent representation is a true representation. By contrast, a phase factor appears in \eqref{33}, showing that the representation of the loop prolongation is itself a cocycle representation. 

The repeated application of \eqref{33} and \eqref{35} gives
\begin{equation}
\left(\hat{U}(\bar{g}_3)\hat{U}(\bar{g_2})\right)\hat{U}(\bar{g}_1)=\hat{U}(\bar{g}_3)\left(\hat{U}(\bar{g}_2)\hat{U}(\bar{g}_1)\right)\label{37}
\end{equation}
This shows that \emph{even though the Galilean line loop $\bar{\cal G}$ is  non-associative, the operators $\hat{U}(\bar{g})$ furnishing its representation 
do associate, a necessary requirement for linear operators in a vector space}. This is a direct outcome of the structure of the associators 
\eqref{27} and composition \eqref{33} in that the difference between the phases on the two sides of \eqref{37} exactly cancels the operator $\hat{U}(A(\bar{g}_3,\bar{g}_2,\bar{g}_1))$ 
representing the corresponding associator. The details of this calculation as well as the construction of the representation \eqref{29} can be found in \cite{klink2}.  In sharp contrast, the operators 
representing symmetry transformations do not associate, and thus are not even defined in the three-cocycle analysis of \cite{jackiw}.

Even though $\bar{\cal G}$ is a loop, all of the physical transformations that correspond to the usual observables form one parameter subgroups. Hence, the observables 
 for a free particle in a non-inertial reference frame can be obtained from \eqref{29} in the usual way.  For the Hamiltonian, we use the definition that it is the generator of time translations, 
i.e., $\hat{H}:=i\left.\frac{d\hat{U}(0,I,\bs{0},b)}{db}\right|_{b=0}$, and obtain from \eqref{29},
\begin{equation}
\hat{H}=\frac{\hat{\bs{P}}^2}{2m}+m\dot{\bs{q}}\cdot\left(\hat{\bs{X}}+\frac{1}{2}\bs{a}_{\bs{q}}\right)\label{39}
\end{equation}
where $\hat{\bs{P}}$ is the momentum operator, defined as the generator of \emph{constant} spatial translations $\bs{a}^{(0)}$, and $\hat{\bs{X}}$ is the position operator canonically conjugated to $\hat{\bs{P}}$, 
$\hat{\bs{X}}:=\frac{i}{m}\nabla_{\bs{q}}$. The first term of \eqref{39} is the usual kinetic energy term and the second term carries the non-inertial effects of the reference frame. As one might expect from classical physics, it is proportional to the inertial mass. However, 
as shown in \cite{klink3},  more general cocycles lead to more complicated expressions for the Hamiltonian, some showing violations of the equivalence principle. 

\noindent\emph{Simulated  magnetic fields from rotating reference frames:}  As a special case, consider the transformation from an inertial reference frame to a rotating reference frame. Then, from \eqref{29} and \eqref{31}, 
\begin{eqnarray}
|\bs{q}\rangle&=&\hat{U}(0,R,\bs{0},0)|\bs{q}^{(0)}\rangle\label{41}\\
\bs{q}&=&R\bs{q}^{0}+\Omega\times\bs{a}_{\bs{q}}\label{43}
\end{eqnarray}
where the superscript $(0)$ indicates the time independence of velocity in an inertial frame. Differentiating \eqref{43}, we obtain 
\begin{equation}
\dot{\bs{q}}=\dot{\Omega}\times\bs{a}_{\bs{q}}+2\Omega\times\bs{q}-\Omega\times(\Omega\times\bs{a}_{\bs{q}})\label{45}
\end{equation}
where the Coriolis and centrifugal terms have emerged naturally. The substitution  of \eqref{45} in the expression for the Hamiltonian 
gives
\begin{equation}
\hat{H}=\hat{A}_0+\frac{1}{2m}\left(\hat{\bs{P}}-\hat{\bs{A}}\right)^2\label{47}
\end{equation}
where 
\begin{eqnarray}
\hat{A}_0&=&-2m\left(\Omega\times(\hat{\bs{X}}+\frac{1}{2}\bs{a}_{\bs{q}})\right)\cdot(\Omega\times\hat{\bs{X}})+m\bs{a}_{\bs{q}}\cdot(\dot{\Omega}\times\hat{\bs{X}})\nonumber\\
\hat{\bs{A}}&=&2m\Omega\times(\hat{\bs{X}}+\frac{1}{2}\bs{a}_{\bs{q}})\label{49}
\end{eqnarray}
Hence, we have shown that the gauge connection that appears in a rotational reference frame can be properly derived from the representations of the Galilean line loop. 
The $\frac{1}{2}\bs{a}_{\bs{q}}$ in $\hat{\bs{X}}+\frac{1}{2}\bs{a}_{\bs{q}}$ is not significant as it results from the choice of the two-cocycle \eqref{9} and may be removed by a suitable phase.  The vector potential $\hat{\bs{A}}$ and the first term of the scalar potential $\hat{A}_0$ are the same as what has been obtained previously 
by the analogy between the Coriolis and Lorentz forces. However, besides being  rigorous, our result holds  for non-constant $\Omega$ as well and in that case there is another term, $m\bs{a}_{\bs{q}}\cdot(\dot{\Omega}\times\hat{\bs{X}})$, 
which would be missed if only rotations were considered, rather than the full Galilean line group.  This suggests a more intricate gauge structure that should be tested by new interferometry experiments with time dependent $\Omega$.

\end{document}